\documentclass[showpacs,eqsecnum,aps]{revtex4}
\usepackage{amsmath}
\usepackage{amssymb}
\usepackage{amsfonts}
\usepackage[dvips]{graphicx,color}
\newcommand{\beq}{\begin{equation}}
\newcommand{\eeq}{\end{equation}}
\newcommand{\beqarray}{\begin{eqnarray}}
\newcommand{\eeqarray}{\end{eqnarray}}

\newcommand{\bLambda}{\mbox{\boldmath$\Lambda$}}

\newcommand{\bphi}{\mbox{\boldmath$\phi$}}

\newcommand{\directsumint}{\sum\mspace{-25mu}\int\,_{\oplus}}

\begin{document}

\title{Mini review of Poincar\'e invariant quantum theory} 

\author{W. N. Polyzou,  Y. Huang}
\affiliation{
Department of Physics and Astronomy, 
The University of Iowa, Iowa City, IA
52242
}

\author{Ch. Elster}
\affiliation{
Institute of Nuclear and Particle Physics  
and Department of Physics and Astronomy,
Ohio University, 
Athens, OH 45701}

\author{W. Gl\"ockle}
\affiliation{
Institut f\"ur theoretische Physik II,
Ruhr-Universit\"at Bochum,
NSF Division of Nuclear Physics
}

\author{ J. Golak,  R. Skibi\'nski, H. Wita{\l}a}
\affiliation{
M. Smoluchowski Institute of Physics, 
Jagiellonian University,
PL-30059 Krak\'ow, Poland
}

\author{H. Kamada} 
\affiliation{ 
Department of Physics, Faculty of Engineering,
Kyushu Institute of Technology,
1-1 Sensuicho Tobata, Kitakyushu 804-8550, Japan
}
\vspace{10mm}

\date{\today}

\begin{abstract}
We review the construction and applications of exactly Poincar\'e
invariant quantum mechanical models of few-degree of freedom systems.
We discuss the construction of dynamical representations of the
Poincar\'e group on few-particle Hilbert spaces, the relation to
quantum field theory, the formulation of cluster properties, and
practical considerations related to the construction of realistic
interactions and the solution of the dynamical equations.  Selected
applications illustrate the utility of this approach.
\end{abstract}


\vspace{10mm}


\maketitle


%

\section{Introduction}

While there is strong evidence that QCD is the theory of the strong
interactions, direct calculations of scattering observables in QCD
with mathematically controlled errors are difficult at some important
energy scales.  These difficulties are particularly significant at the
few-GeV scale, where perturbative methods are not applicable.  This is
an interesting energy scale because it is the scale where sensitivity
to sub-nuclear degrees of freedom is expected to begin.  Mathematical
models that are motivated by QCD may provide useful insight into the
dynamics at these energy scales.

Poincar\'e invariant quantum mechanics is one of a number of
approaches that can be used to model systems of strongly interacting
particles at the few GeV energy scale.  At the simplest level it is
quantum mechanics with an underlying Poincar\'e symmetry.  While
Poincar\'e invariant quantum mechanics can be treated as a
phenomenology that is independent of QCD, it can also be
related to QCD. Poincar\'e invariant quantum mechanics has proved to
be useful in applications, but there are no textbook treatments of the
subject.

Historically, Poincar\'e invariant quantum mechanics was first
articulated by Wigner \cite{Wigner:1939cj}, who pointed out that a
necessary and sufficient condition for a quantum theory to be
relativistically invariant is the existence of a unitary ray
representation of the Poincar\'e group on the quantum mechanical
Hilbert space.  
Wigner's work did not have a significant impact on applications of
quantum field theory, but it directly motivated attempts to 
provide an axiomatic
\cite{Wightman:1980,Haag:1963dh,Osterwalder:1973dx} foundation for
quantum field theory.  These axioms provide a Hilbert space
formulation of quantum field theory that can be directly related to
Poincar\'e invariant quantum mechanics.

Dirac \cite{Dirac:1949cp} studied the problem of constructing the
Poincar\'e Lie algebra for systems of interacting particles.  He
observed that the presence of interactions in the Hamiltonian implied
that at most a sub-algebra of the Poincar\'e Lie algebra could be free
of interactions.  He identified the three largest sub-algebras, and
classified dynamical models according to which sub-algebra remained
free of interactions.  Dirac used the terms instant, point, and
front-forms of dynamics to label the different kinematic sub-algebras.
Bakamjian and Thomas \cite{Bakamjian:1953kh} provided the first
construction of the full Poincar\'e Lie algebra for a system of two
interacting particles in Dirac's instant-form of the dynamics.
Coester \cite{Coester:1965zz} generalized Bakamjian and Thomas'
construction to systems of three interacting particles.  His
construction also led to a $S$ matrix that satisfied spacelike cluster
properties.  Sokolov \cite{Sokolov:1977} provided a complete
construction of the Poincar\'e Lie Algebra for a system of $N$
interacting particles in Dirac's point-form of the dynamics that was
consistent with a stronger form of spacelike cluster properties, where
the Poincar\'e generators satisfy cluster properties.  This stronger
form of cluster properties provides a simple relation between the few
and many-body systems that is difficult to realize in theories
satisfying only $S$-matrix cluster properties.  Coester and Polyzou 
\cite{Coester:1982vt} provided the complete solution for
systems of $N$-particles in all three of Dirac's forms of the dynamics
satisfying the strong form of cluster properties.  A more general
construction based on only group representations, that has Dirac's
form of dynamics as special cases, was given in
\cite{Polyzou:19892b,Polyzou:2003dt}.  The subject was reviewed
by Keister and Polyzou in \cite{Keister:1991sb}.

There have been many applications of Poincar\'e invariant quantum
mechanics in all three of Dirac's forms of dynamics. 
The earliest applications involved the study of electromagnetic 
probes on mesons, nucleons, and nuclei.  Some of the relevant 
papers are 
\cite{Bakker:1979eg,Grach:1983hd,Chung:1988my,Chung:1988mu,Cardarelli:1995dc,Polyzou:1996fb,Krutov:1997wu,Allen:2000ge,Wagenbrunn:2000es,JuliaDiaz:2003gq,Sengbusch:2004sf,Coester:2005cv,Huang:2008jd,Arrington:2008zh,Desplanques:2009kj}. 
The  first three-nucleon bound state 
calculation using this framework was performed
by Gl\"ockle, Coester and Lee \cite{Glockle:1986zz}.
Calculations of the triton binding energy with realistic interactions have 
been performed recently \cite{Kamada:2008xc}.
Applications to nuclear reactions appear in  
\cite{Fuda:1995zz,Lin:2007ck,Lin:2007kg,Lin:2008sy,Witala:2008va,Witala:2009zzb}
 which include reactions with particle production
\cite{Fuda:2009zz}.

This mini-review is limited to theories that are formulated by
constructing exact unitary representations of the Poincar\'e group on
few-particle Hilbert spaces.  There are many other approaches to
relativistic quantum mechanics that have been successfully applied at
the few GeV scale.  Each one emphasizes different desirable features
of the full field theory, however when the number of degrees of
freedom is limited, it is impossible to satisfy all of the axioms of
the underlying field theory.  
Our preference for using Poincar\'e invariant quantum mechanics is
based on three observations: (1) many computational methods
successfully used in non-relativistic quantum mechanics can be
directly applied in Poincar\'e invariant quantum
mechanics, (2) the theories involve a finite number of degrees of
freedom, allowing exact numerical calculations of model predictions, 
(3) the theories share most of the axiomatic properties of quantum field
theory and there is a direct relation to the Hilbert space formulation
of field theory.  The fundamental property of the quantum field theory
that is given up in order to have a theory of a finite number of
degrees of freedom is microscopic locality.  The justification for
this choice is that microscopic locality is not an experimentally
testable property since probing a system at arbitrarily short distance
scales requires arbitrarily large energy transfers.  In addition,
Poincar\'e invariant quantum mechanics does not have a large enough
algebra of observables to localize particles in arbitrarily small
spacetime regions.  One manifestation of this is the absence of a
reasonable position operator \cite{Newton:1949cq} in relativistic
quantum theories of a finite number of degrees of freedom. 

In the next section we discuss the construction of representations of
single-particle Hilbert spaces.  In section \ref{sec3} we discuss
irreducible representations of the Poincar\'e group that act on the
single-particle Hilbert spaces.  In section
 \ref{sec4} we construct a dynamical representation of the Poincar\'e group
by adding interactions to the mass Casimir operator of a
non-interacting irreducible representation constructed from tensor
products of single particle representations. The strong and weak form
of cluster properties are discussed in section \ref{sec5}.  The formulation
of the three-body problem is discussed in section \ref{sec6}.  The relation
to quantum field theory is discussed in section \ref{sec7}.  Selected
few-nucleon applications are discussed in section \ref{sec8}.

\section{Particles, Hilbert spaces and irreducible representations}

Experiments measure observables that describe the state of free
particles by considering how the particles interact with classical
electromagnetic fields.  A complete experiment measures the linear
momentum and spin state of each initial and final particle.  There is
a natural connection with these single-particle observables and
irreducible representations of the Poincar\'e group.  The Poincar\'e
group has ten infinitesimal generators.  These Hermitian operators
include the Hamiltonian which generates time translations, the linear
momentum operators which generate space translations, the angular
momentum operators which generate rotations, and the rotationless
boost generators which generate transformations that change the
momentum of the particle.  From these ten infinitesimal generators it
is possible to construct two Casimir invariants, four independent
commuting Hermitian observables and four conjugate operators.  The
Casimir invariants fix the mass and spin of the particle.  Eigenvalues
of the commuting observables label the states of the particle, and the
conjugate operators determine the spectrum of the commuting
observables and thus the allowed states of the particle.

For a standard description of a particle, the commuting observables
can be taken to be the three components of the linear momentum, and a
component of a spin operator.  The spectrum of the momentum is
$\mathbb{R}^3$, while the spectrum of a component of the spin vector
takes on discrete values in integer steps from $-j$ to $j$.  In this
case the Hilbert space is
\beq 
{\cal H}_{mj} = L^2 (\mathbb{R}^3)\otimes \mathbb{C}^{2j+1} .
\label{b.1}
\eeq
Single-particle states are represented by wave functions, 
$ 
\psi (\mathbf{p},\mu ) = \langle (m,j) \mathbf{p},\mu  \vert \psi \rangle .
$
The relations of the operators
$m,j^2,\mathbf{p}$ and $\mathbf{j}\cdot \hat{\mathbf{z}}$ 
to 
the Poincar\'e Lie Algebra determines a unitary representation, 
$U_1(\Lambda ,a)$, of Poincar\'e group on ${\cal H}_{mj}$:
\beq
\langle (m,j) \mathbf{p},\mu  \vert U_1(\Lambda ,a) \vert \psi \rangle =
\int \sum_{\mu'=-j}^j 
{\cal D}^{m,j}_{\mathbf{p},\mu;\mathbf{p}',\mu'}[\Lambda ,a]
d\mathbf{p}'  \psi (\mathbf{p}',\mu' ) = \psi' (\mathbf{p},\mu )
\label{b.3}
\eeq
where the Poincar\'e group Wigner function is
\[
{\cal D}^{m,j}_{\mathbf{p},\mu;\mathbf{p}',\mu'}[\Lambda ,a] : =
\langle (m,j) \mathbf{p},\mu  \vert U_1(\Lambda ,a) \vert 
(m,j) \mathbf{p}',\mu'  \rangle =
\]
\beq
\delta (\mathbf{p} - \bLambda p' )\sqrt{{\omega_m (\mathbf{p}) \over
\omega_m (\mathbf{p}')}}
e^{i p\cdot a} D^j_{\mu \mu'}[R_{wc} (\Lambda,p)]
\label{b.4}
\eeq
and $R_{wc} (\Lambda,p)$ is a Wigner rotation.

Because a sequence of Lorentz boosts that start and end at the rest
frame generally define a rotation, in order to obtain an unambiguous
definition of a spin vector for all values of the particle's momentum,
it is necessary to define a standard way to measure a spin observable.
The above representation implicitly defines the spin projection by its
value in the particle's rest frame after the particle is transformed
to the rest frame with a rotationless Lorentz transformation.  This is
one of an infinite number of possible choices of spin observables.
This choice is consistent with the ``canonical'' spin that appears in
standard Dirac $u$ and $v$ spinors.  Different spin observables are
related by momentum-dependent rotations that lead to different couplings to
the electromagnetic field.  This ensures that measurable
physical quantities are independent of the observables used to label
single particle states.

These single-particle representations are irreducible, and all
positive-mass positive-energy irreducible representations of the
Poincar\'e group can be put in this general form.  These irreducible
representations will be important in formulating dynamical models.  In
general, any unitary representation of the Poincar\'e group can be
decomposed into a direct sum or direct integral (for continuous mass
eigenvalues) of irreducible representations.  We will build the
dynamical unitary representation of the Poincar\'e group out of the
non-interacting irreducible representations.

\section{Poincar\'e group Wigner functions 
and kinematic subgroups}

In the previous section we represented single-particle wave functions
in the basis of generalized eigenstates
$\vert (m,j), \mathbf{p} , \mu \rangle$ .

The state of the particle could be also alternatively determined by
measuring the particles' four velocity, 
$v^{\mu}= (\sqrt{1-\mathbf{v}\cdot \mathbf{v}},\mathbf{v})$,
and spin projection: 
\beq
\vert (m,j), \mathbf{v} , \mu \rangle = 
\vert (m,j), \mathbf{p}(\mathbf{v},m)  , \mu \rangle m^{3/2},
\label{c.2}
\eeq
or the light-front components of the four momentum 
$p^+= \sqrt{m^2 + \mathbf{p}^2}+ \mathbf{p}\cdot \hat{\mathbf{z}}$,
$\mathbf{p}_{\perp}= (\mathbf{p}\cdot \hat{\mathbf{x}},
\mathbf{p}\cdot \hat{\mathbf{y}})$,
and light-front spin projection:
\beq 
\vert (m,j), p^+ , \mathbf{p}_{\perp} , \mu \rangle = \sum_{\mu'=-j}^j 
\vert (m,j), \mathbf{p}(p^+,\mathbf{p}_{\perp},m)  , \mu' \rangle 
\sqrt{{\omega_m (\mathbf{p})  \over p^+ }}
D^j_{\mu' \mu} [B_c^{-1}(p) B_f(p)]
\label{c.3}
\eeq
where $B_c^{-1}(p) B_f(p)$
is a Melosh rotation \cite{Melosh:1974cu},
defined by a light-front-preserving boost
followed by the inverse of a rotationless boost.  The different basis
choices are related to the basis $\vert (m,j), \mathbf{p} , \mu \rangle$
by the unitary
transformations in (\ref{c.2}) and (\ref{c.3}).  The light-front
preserving boosts have the desirable property that they form a group,
which means that there are no Winger rotations for any sequence of
light-front preserving boosts that start and end in the rest frame;
the price paid for this is that the Wigner rotation of a pure rotation
is not equal to the rotation.  

The Poincar\'e group Wigner functions depend on the choice of basis.
The Wigner functions 
\beq
{\cal D}^{m,j}_{\mathbf{v},\mu;\mathbf{v}',\mu'}[\Lambda ,a] : =
\langle (m,j) \mathbf{v},\mu  \vert U_1(\Lambda ,a) \vert 
(m,j) \mathbf{v}',\mu'  \rangle 
\label{c.6}
\eeq
\beq
{\cal D}^{m,j}_{p^+,\mathbf{p}_{\perp},\mu;
p^{\prime +},\mathbf{p}'_{\perp},\mu'}[\Lambda ,a] : =
\langle (m,j) p^+,\mathbf{p}_{\perp},\mu  \vert U_1(\Lambda ,a) \vert 
(m,j) p^{\prime +},\mathbf{p}_{\perp}',\mu'  \rangle
\label{c.7}
\eeq
are related to the Wigner function (\ref{b.4}) 
by the unitary transformations (\ref{c.2})
and (\ref{c.3}).

While the concept of a kinematic subgroup does not make sense for a
single particle, the kinematic subgroup for an instant-form dynamics
is the subgroup of the Poincar\'e group that leaves the Wigner
function (\ref{b.4}) independent of mass; the kinematic subgroup for a
point-form dynamics is the subgroup of the Poincar\'e group that
leaves the Wigner function (\ref{c.6}) independent of mass; the
kinematic subgroup for a front-form dynamics is the subgroup of the
Poincar\'e group that leaves the Wigner function (\ref{c.7})
independent of mass.  Different mass-independent subgroups appear in
different irreducible bases because the unitary transformations
relating the irreducible bases (\ref{c.2}) and (\ref{c.3}) to the
basis $\vert (m,j) \mathbf{p},\mu \rangle$ depend on the particles'
mass.  These mass-independent subgroups become kinematic subgroups in
dynamical models because the mass acquires an interaction while the
other operators used to construct dynamical irreducible bases remain
interaction free.  More generally, it is possible to define perfectly
good single-particle bases where the identity is the only subgroup
where the corresponding Wigner function is independent of mass.

\section{Two-body models - Clebsch-Gordan 
coefficients}

The two-body Hilbert space is a tensor product of two 
single-particle Hilbert spaces,
${\cal H} = {\cal H}_{m_1j_1} \otimes {\cal H}_{m_2j_2} .$
The non-interacting representation of the Poincar\'e group
on ${\cal H}$ is the tensor product of two single-particle
(irreducible) representations of the Poincar\'e group,
$U_0 (\Lambda,a ) := U_1 (\Lambda,a ) \otimes U_2 (\Lambda,a )$ .
While the single-particle representations of the Poincar\'e group are
irreducible, their tensor product is reducible.  Formally the tensor
product representation can be expressed as a direct integral of
irreducible representations,
\beq
U_0 (\Lambda,a ) = {\directsumint}_{jls} dm U_{0,m,j,l,s} (\Lambda ,a), 
\label{d.3}
\eeq
where $U_{0,m,j, l,s} (\Lambda ,a)$ are mass $m$ spin $j$ irreducible
representations of the Poincar\'e group.  The quantum numbers $l$ and
$s$ are invariant degeneracy parameters that distinguish multiple
copies of the irreducible representations with the same $m$ and $j$.  
They have the same quantum numbers as the spin and orbital angular momentum.
The mass $m$ is the two-particle invariant mass that has a continuous 
spectrum starting from $m_1+ m_2$.  The Poincar\'e group Clebsch-Gordan 
coefficients relate the tensor product representation to the direct integral 
of irreducible representations and satisfy
\[
\sum \int {\cal D}^{m,j}_{\mathbf{p},\mu;\mathbf{p}',\mu'}[\Lambda ,a] d\mathbf{p}'
\langle (m,j,l,s), \mathbf{p}' , \mu' \vert
(m_1,j_1), \mathbf{p}_1 , \mu_1 ;
(m_2,j_2), \mathbf{p}_2 , \mu_2 \rangle  =
\]
\[
\sum \int
\langle (m,j,l,s), \mathbf{p} , \mu \vert
(m_1,j_1), \mathbf{p}_1' , \mu_1' ;
(m_2,j_2), \mathbf{p}_2' , \mu_2' \rangle d\mathbf{p}_1'\mathbf{p}_2'
\times
\]
\beq
{\cal D}^{m_1,j_1}_{\mathbf{p}_1',\mu_1';\mathbf{p}_1,\mu_1}
[\Lambda ,a] 
{\cal D}^{m_2,j_2}_{\mathbf{p}_2',\mu_2';\mathbf{p}_2,\mu_2}[\Lambda ,a] .
\label{d.2}
\eeq
The Clebsch-Gordan coefficients, 
$
\langle (m,j,d), \mathbf{p} , \mu \vert
(m_1,j_1), \mathbf{p}_1 , \mu_1 ;
(m_2,j_2), \mathbf{p}_2 , \mu_2 \rangle
$, $d:=\{l,s\}$,
are basis-dependent and are known in all three of the
representations 
(\cite{Joos:1962,Coester:1965zz,Moussa:1965,Keister:1991sb}). 

The two-body irreducible basis states look similar to relative and
center of mass variables in non-relativistic quantum mechanics; but
they differ in the structure of the Poincar\'e group Clebsch-Gordan
coefficients, which contain momentum-dependent spin rotation functions
and non-trivial kinematic factors that ensure unitarity.

The basis states $\{\vert (m,j,l,s), \mathbf{p}
, \mu \rangle \}$ transform irreducibly under $U_0(\Lambda ,a)$:
\beq
U_0 (\Lambda ,a) 
\vert (m,j,d), \mathbf{p} , \mu \rangle
=
\sum' \int d \mathbf{p}' 
\vert (m,j,d), \mathbf{p}' , \mu' \rangle 
{\cal D}^{m,j}_{\mathbf{p}',\mu';\mathbf{p},\mu}[\Lambda ,a] .
\label{d.5}
\eeq

While (\ref{d.5}) is {\it not} the dynamical representation of the
Poincar\'e group, by working in this non-interacting irreducible basis
it is possible to construct dynamical representations by adding an
interaction $v$, which in this basis has a kernel of the form
\beq
\langle (m',j',d'), \mathbf{p}' , \mu' \vert v
\vert (m,j,d), \mathbf{p} , \mu \rangle
=
\delta_{j'j}\delta_{\mu' \mu}\delta (\mathbf{p}'-\mathbf{p})
\langle m' ,d' \Vert v^j
\Vert m , d  \rangle , 
\label{d.6}
\eeq
to the non-interacting two-body mass operator.
This interaction has the same form as a typical Galilean invariant 
non-relativistic interaction if we replace $m= \sqrt{m_1^2 + {k}^2} + 
\sqrt{m_2^2 + {k}^2}$ by $k$ and $d=\{l,s\}$ by $(l,s)$.

We define the dynamical mass operator 
$ M:= \sqrt{m_1^2 + {k}^2} + 
\sqrt{m_2^2 + {k}^2} + v$ .
Simultaneous eigenstates of $M$, $\mathbf{p}$, $\mathbf{j}^2$
and $\mathbf{j}  \cdot \hat{\mathbf{z}}$ 
can be constructed by diagonalizing $M$ in the irreducible 
non-interacting basis.
These eigenfunctions 
have the form  
\beq
\langle ( k',j',l',s'), \mathbf{p}' , \mu' \vert
(\lambda,j), \mathbf{p} , \mu \rangle =
\delta_{j'j}\delta_{\mu' \mu}\delta (\mathbf{p}'-\mathbf{p})
\phi_{\lambda ,j}(\mathbf{k}^2,l,s)
\label{d.9}
\eeq
where the wave function, $\phi_{\lambda ,j}(k^2,l,s)$, is the solution of the
mass eigenvalue problem with eigenvalue $\lambda$:
\[
(\lambda - \sqrt{m_1^2 + k^2} - 
\sqrt{m_2^2 + k^2} ) \phi_{\lambda ,j}(k,l,s) =
\]
\beq
\int_0^{\infty}  k^{\prime 2} dk' \sum_{s'} \sum_{l'=\vert j-s \vert}^{j+s} 
\langle k,l,s \vert V^j
\vert k',l',s'  \rangle \phi_{\lambda ,j}(k',l',s').
\label{d.10}
\eeq

The dynamical unitary representation of the Poincar\'e group 
is defined on this complete set of states,  
$\vert
(\lambda,j), \mathbf{p} , \mu \rangle$ by
\beq
U(\Lambda ,a)  \vert
(\lambda,j), \mathbf{p} , \mu \rangle
 =
\sum_{\mu'=-j}^j \int d\mathbf{p}' \vert
(\lambda,j), \mathbf{p}' , \mu' \rangle  
{\cal D}^{\lambda,j}_{\mathbf{p}',\mu';\mathbf{p},\mu}[\Lambda ,a] .
\label{d.11}
\eeq
The relevant dynamical feature is that the Poincar\'e group Wigner
function now depends on the eigenvalue $\lambda$ of the dynamical mass
operator, which requires solving (\ref{d.10}).

Because of the choice of basis, the Poincar\'e group Wigner function $
{\cal D}^{\lambda,j}_{\mathbf{p}',\mu';\mathbf{p},\mu}[\Lambda ,a] $
has the same structure as the Wigner function (\ref{b.4}) and thus has
the property that when $(\Lambda,a)$ is in the three-dimensional
Euclidean subgroup, it is independent of the mass eigenvalue
$\lambda$, which means that for this dynamical model the kinematic
subgroup is dictated by the {\it choice of representation used to define
the irreducible basis}.

Even though the dynamics has a non-trivial interaction dependence, 
it is only necessary to solve (\ref{d.10}), which is analogous to 
solving the center of mass Schr\"odinger equation in the 
non-relativistic case.

This construction can be repeated using different irreducible bases,
such as (\ref{c.2}) or (\ref{c.3}), where the Wigner functions have
different mass-independent symmetry groups.  For these bases if we choose
to use the interactions 
\beq 
\langle (k',j',l',s'), \mathbf{v}' , \mu'
\vert v_{point} \vert (k,j,l,s), {v} , \mu \rangle =
\delta_{j'j}\delta_{\mu' \mu}\delta (\mathbf{v}'-\mathbf{v}) \langle
k',l',s' \Vert v^j \Vert k,l,s \rangle 
\label{d.12}
\eeq 
\[
\langle
(k',j',l',s'), p^{\prime +},\mathbf{p}'_{\perp} , \mu' \vert v_{front} \vert
(k,j,l,s), p^{+},\mathbf{p}_{\perp}, \mu \rangle =
\]
\beq
\delta_{j'j}\delta_{\mu' \mu}\delta
(\mathbf{p}_{\perp}'-\mathbf{p}_{\perp}) \delta (p^{\prime +} -p^+)
\langle k',l',s' \Vert v^j \Vert k,l,s \rangle, 
\label{d.13}
\eeq 
where the reduced kernels $\langle k',l',s' \Vert v^j \Vert k,l,s \rangle$
are the same in (\ref{d.6}), (\ref{d.12}), and (\ref{d.13}) 
in the bases (\ref{b.3}), (\ref{c.2}), and (\ref{c.3}), respectively, 
and construct dynamical
eigenstates of the form
\beq 
\vert (\lambda,j), \mathbf{v} , \mu \rangle, \qquad \vert
(\lambda,j), p^{+},\mathbf{p}_{\perp}, \mu \rangle  ~,
\label{d.14}
\eeq 
then equation (\ref{d.10}) still determines the binding energy and
scattering phase shifts.  It follows that the
resulting two-body models have the same bound-state and scattering
observables, however each of the resulting unitary representations of
the Poincar\'e group has a different kinematic subgroup.  The
dynamical irreducible eigenstates transform like $\vert (\lambda,j),
\mathbf{p} , \mu \rangle$ with the Wigner function (\ref{d.11})
replaced by (\ref{c.6}) or (\ref{c.7}) where $m$ is replaced by
$\lambda$.  This makes these unitary transformations dynamical.

The mass operators and interactions, $v$, $v_{point}$ and $v_{front}$
are distinct operators, but the three representations are
related by unitary transformation that leave the binding energies and
scattering observables unchanged.  The dynamical calculations are
identical in all three cases and are given by solving (\ref{d.10}).
This shows that dynamical models with different kinematic subgroups
are equivalent and cannot be distinguished on the basis of any
experimental observations.

\section{Cluster properties - Ekstein's theorem}

An important feature of non-relativistic quantum mechanics is that the
same interactions appear in the few and many-body problems.
Specifically, the Hamiltonian becomes a sum of subsystem Hamiltonians
when the short-ranged interactions between particles in different
subsystems are turned off.  In the relativistic case the corresponding
requirement is that the unitary time-translation group breaks up into
a tensor product of subsystem groups when the system is asymptotically
separated into independent subsystems.  We call this the strong form
of cluster properties.

The observable requirement is that the $S$-matrix clusters.  We call
this the weak form of cluster properties because it follows from the
strong form of cluster properties, however because different
Hamiltonians can have the same $S$-matrix, the weak form of cluster
properties does not imply that the same interactions appear in the few
and many-body Hamiltonians.  Because of this, in order to maintain a
simple relation between the few and many-body problem, we require that
Poincar\'e invariant quantum theories satisfy the strong form of
cluster properties.

A theorem of Ekstein \cite{Ekstein:1960} provides necessary and
sufficient conditions for two short-ranged interactions to give the
same $S$ matrix.  The requirement is that the Hamiltonians are related
by a unitary transformation $A$ satisfying the asymptotic condition
\beq
\lim_{t \to \pm \infty} \Vert (I-A) U_0(t) \vert \psi \rangle \Vert =0
\label{e.1}
\eeq
where $U_0(t)$ is the non-interacting time translation operator.  We
refer to unitary transformations with this property as scattering
equivalences.  It is important that this condition be satisfied for
both time limits; to appreciate the relevance of this condition
consider two Hamiltonians with different repulsive potentials. Because
these Hamiltonians have the same spectrum and multiplicities they are
related by a unitary transformation, however the derived S-matrices
may have different phase shifts.  The phase shifts differ if and only
if two time limits do not agree.

Scattering equivalences that preserve weak cluster properties but not
strong cluster properties exist and are the key to restoring the
strong form of cluster properties in Poincar\'e invariant quantum
theory.  The strategy is illustrated in the formulation of the 
three-body problem in the next section.

\section{Three-body problem}

The strong form of cluster properties implies that given a 
set of dynamical two-body generators, the three-body generators
necessarily can be expressed as sums of one, two and three-body 
operators
\beq
H = H_1 + H_2 + H_3 + H_{12} + H_{23} + H_{31} + H_{123} 
\label{f.1}
\eeq
\beq
\mathbf{P}  = \mathbf{P}_1 + \mathbf{P}_2 + \mathbf{P}_3  + 
\mathbf{P}_{12} + \mathbf{P}_{23} + \mathbf{P}_{31} + \mathbf{P}_{123} 
\label{f.2}
\eeq
\beq
\mathbf{J}  = \mathbf{J}_1 + \mathbf{J}_2 + \mathbf{J}_3  + 
\mathbf{J}_{12} + \mathbf{J}_{23} + \mathbf{J}_{31} + \mathbf{J}_{123} 
\label{f.3}
\eeq 
\beq
\mathbf{K}  = \mathbf{K}_1 + \mathbf{K}_2 + \mathbf{K}_3  + 
\mathbf{K}_{12} + \mathbf{K}_{23} + \mathbf{K}_{31} + \mathbf{K}_{123} . 
\label{f.4}
\eeq 
The one and two-body operators in (\ref{f.1}-\ref{f.4}) 
are the same operators that appear in the 
two-body problems, while the three-body operators,
$H_{123},\mathbf{P}_{123}, \mathbf{J}_{123}$ and $\mathbf{K}_{123}$
are the only new ingredients in the three-particle generators.

It is easy to show that if the generators have this form it is impossible 
to satisfy the Poincar\'e commutation relations if all of the 
three-body operators vanish.  However, although the commutation 
relations put non-linear constraints on these operators, it will become 
clear that the solutions are not unique.

To avoid solving the non-linear problem of satisfying the commutation
relations, it is more productive to start by first satisfying the
commutation relations at the expense of strong cluster properties.
This can be done by applying the method of section \ref{sec4} directly to the
three-body problem.  This involves adding suitable interactions to the
non-interacting invariant three-body mass operator.  

To begin the construction we consider a three-body system where only
one pair of particles interact.  The relevant basis is a
non-interacting three-body irreducible representation of the
Poincar\'e group.  It is constructed by successive pairwise coupling
using the Poincar\'e group Clebsch-Gordan coefficients.  If we assume
that particles one and two are the interacting pair then preferred order
of coupling would be $(12)\to((12)(3))$:
\beq
\vert \mathbf{p}_1, \mu_1 \rangle \otimes
\vert \mathbf{p}_2, \mu_2 \rangle \to 
\vert (k_{12},l_{12},s_{12},j_{12})\mathbf{p}_{12}, \mu_{12} \rangle
\label{f.5}
\eeq
\[
\vert (k_{12},l_{12},s_{12},j_{12})\mathbf{p}_{12}, \mu_{12} \rangle
\otimes \vert \mathbf{p}_3, \mu_3 \rangle \to 
\]
\beq
\vert (q,L_{(12)(3)},S_{(12)(3)}J_{(12)(3)},k_{12},l_{12},s_{12},j_{12})
\mathbf{p}, \mu \rangle .
\label{f.6}
\eeq
We introduce the following shorthand notation for the basis states in these
equations.  We write (\ref{f.5}) as $\vert 1 \otimes 2 \rangle \to
\vert (12) \rangle$ and (\ref{f.6}) as $\vert (12)\otimes 3 \rangle
\to \vert (12)(3) \rangle$. Using this notation we define two different
embeddings of the two-body interaction in the three-body Hilbert space
using the two representation in (\ref{f.6}):
\[
\langle (12)'\otimes 3'  \vert
v_{12\otimes}
\vert (12)\otimes 3  \rangle =
\]
\beq
\langle k_{12}',l_{12}',s_{12}' \Vert v^j \Vert 
k_{12},l_{12},s_{12},\rangle
\delta (\mathbf{p}_{12}' -\mathbf{p}_{12} )
\delta (\mathbf{p}_{3}' -\mathbf{p}_{3} )  
\delta_{j_{12}'j_{12}}\delta_{j_3'j_3}
\label{f.7}
\eeq
and
\[
\langle (12)'(3)' \vert {v}_{12} \vert (12)(3)\rangle =
\]
\[
\langle k_{12}',l_{12}',s_{12}' \Vert v^j \Vert 
k_{12},l_{12},s_{12},\rangle
\delta (\mathbf{p}' -\mathbf{p} ) \times
\]
\beq
{\delta (q'-q ) \over q^2}  
\delta_{j_{(12)(3)}'j_{(12)(3)}}\delta_{j_{12}'j_{12}}
\delta_{L_{(12)(3)}'L_{(12)(3)}}\delta_{S_{(12)(3)}'S_{(12)(3)}}
\delta_{\mu'\mu}
\label{f.8}
\eeq
where the reduced kernel, $\langle k_{12}',l_{12}',s_{12}' \Vert v^j \Vert
k_{12},l_{12},s_{12},\rangle$, is identical in (\ref{f.7}) and
(\ref{f.8}).  These expressions define different interactions, $(v_{12\otimes}
\not= {v}_{12})$.

We use these two interactions to define two different $2+1$-body 
mass operators $M_{(12)\otimes (3)}$ and ${M}_{(12)(3)}$ defined by 
\beq
M_{(12)\otimes (3)} := 
\sqrt{
(\sqrt{(\sqrt{m_1^2 + k_{12}^2} + 
\sqrt{m_2^2 + k_{12}^2} +v_{12\otimes})^2 + \mathbf{p}_{12}^2})+ 
 \sqrt{m^2 + \mathbf{p}_3^2})^2 - \mathbf{p}^2)} 
\label{f.9}
\eeq
\beq
{M}_{(12)(3)}:=\sqrt{ 
(\sqrt{m_1^2 + k_{12}^2} + 
\sqrt{m_2^2 + k_{12}^2} + {v}_{12})^2
+q^2} + \sqrt{ {m}_{3}^2+q^2} .  
\label{f.10}
\eeq
Because of the invariance principle
\cite{kato:1966,Chandler:1976,baumgartl:1983} the $S$-matrix can be
computed by replacing the Hamiltonian by the mass operator (this is
equivalent to evaluating the $S$-matrix in the three-body rest frame)
in the standard time-dependent representation of the scattering
operator.

$M_{(12)\otimes (3)}$ is the mass operator of the tensor product of a two-body
representation involving particles one and two and a spectator
representation of the Poincar\'e group associated with particle three,
$U_{12}(\Lambda ,a) \otimes U_3 (\Lambda,a)$.
By construction it is consistent with the strong form of cluster
properties.  The mass operator ${M}_{(12)(3)}$ commutes with the
three-body spin and commutes with and is independent of the total three-body
momentum and $z$-component of the three-body canonical spin.  Simultaneous
eigenstates of ${M}_{(12)(3)},\mathbf{p},j^2,j_z$ are complete and transform
irreducible with respect to the Poincar\'e group.  This defines a
dynamical unitary representation of the Poincar\'e group, $
U_{(12)(3)}(\Lambda ,a)$, on the 
three-body Hilbert space following the construction of section \ref{sec4}.

The scattering operators associated with both of these
operators are related by 
\[
\langle (12)\otimes(3) \vert S_{(12)\otimes (3)} \vert (12)\otimes(3) \rangle=
\]
\beq
\langle k_{12}',l_{12}',s_{12}' \Vert S^j \Vert 
k_{12},l_{12},s_{12},\rangle
\delta (\mathbf{p}_{12}' -\mathbf{p}_{12} )
\delta (\mathbf{p}_{3}' -\mathbf{p}_{3} )  
\delta_{j_{12}'j_{12}}\delta_{j_3'j_3}
\label{f.11}
\eeq
and
\[
\langle (12)(3) \vert S_{(12)(3)} \vert (12 )(3) \rangle =
\langle k_{12}',l_{12}',s_{12}' \Vert S^j \Vert 
k_{12},l_{12},s_{12},\rangle
\delta (\mathbf{p}' -\mathbf{p} ) \times
\]
\beq
{\delta (q'-q ) \over q^2}  
\delta_{j_{(12)(3)}'j_{(12)(3)}}\delta_{j_{12}'j_{12}}
\delta_{L_{(12)(3)}'L_{(12)(3)}}\delta_{S_{(12)(3)}'S_{(12)(3)}}
\delta_{\mu'\mu}
\label{f.12}
\eeq
where the reduced two-body kernels $\langle k_{12}',l_{12}',s_{12}'
\Vert S^j \Vert k_{12},l_{12},s_{12},\rangle$ are identical. 
Because the delta functions become equivalent when they
are evaluated on shell, the $S$ matrices in both representations are
identical.  Ekstein's theorem implies the scattering equivalence
$A_{(12)(3)}U_{12}(\Lambda ,a) \otimes 
U_3 (\Lambda ,a) A_{(12)(3)}^{\dagger} = U_{(12)(3)} (\Lambda ,a)$.

To construct a dynamical representation of the Poincar\'e group 
with all three particles interacting we first construct the mass 
operator
\beq
{M}= {M}_{(12)(3)}+ {M}_{(23)(1)}+
{M}_{(31)}-2M_0 .
\label{f.13}
\eeq
Because each term in (\ref{f.13}) commutes with $\mathbf{p},j^2 ,j_z$,
and is independent of $\mathbf{p}$ and $j_z$ it follows that
simultaneous eigenstates of ${M},\mathbf{p},j^2 ,j_z$ are complete
and transform irreducibly, thus defining a dynamical unitary
representation, ${U}(\Lambda ,a)$, of the Poincar\'e group on the
three-nucleon Hilbert space. 

Because each of the $2+1$ mass operators in (\ref{f.13}) is scattering 
equivalent to $2+1$ mass operators associated with a tensor product 
representation, it follows that ${M}$ can be expressed as
\[
{M} =
A_{(12)(3)} {M}_{(12)\otimes(3)} A_{(12)(3)}^{\dagger} +
A_{(23)(1)} {M}_{(23)\otimes(1)} A_{(23)(1)}^{\dagger} +
\]
\beq
A_{(31)(2)} {M}_{(31)\otimes(2)} A_{(31)(2)}^{\dagger} 
- 2 M_0 .
\label{f.13a}
\eeq
From this representation it follows that when interaction between 
the $i^{th}$ particle and the other two particles are turned off that 
\beq
{U}(\Lambda ,a) \to 
A_{(jk)(i)} {U}_{(jk)}(\Lambda ,a)\otimes
{U}_{i}(\Lambda ,a) A_{(jk)(i)}^{\dagger}
\label{f.14}
\eeq
which formally violates the strong form of cluster properties.

The strong form of cluster properties can be restored by 
transforming ${U}(\Lambda ,a)$ with the product
$A^{\dagger} =A^{\dagger}_{(12)(3)}A^{\dagger}_{(31)(2)}A^{\dagger}_{(23)(1)}$.
Because products of scattering equivalences are scattering equivalences,
this does not change the three-body $S$ matrix.  This transformation 
also restores strong-cluster properties, because  $A^{\dagger}\to 
A_{(jk)(i)}^{\dagger}$ when the interactions between particle $i$ and the 
other two particles are turned off, canceling off the extra
unitary transformations in (\ref{f.14}).  The undesirable feature of $A$
is that the individual $A_{(jk)(i)}$'s do not commute,
so it introduces an exchange asymmetry that does not affect the 
$S$-matrix.  The exchange symmetry can be manifestly restored 
by replacing the product of the $A_{(jk)(i)}$'s     
by a symmetrized product, such as:
\beq
A := e^{\ln (A_{(12)(3)})+\ln (A_{(23)(1)})+\ln (A_{(31)(2)})}
\label{f.15}
\eeq
\beq
{U}_\otimes(\Lambda ,a) = A^{\dagger} {U}(\Lambda ,a) A .
\label{f.16}
\eeq 
Equation (\ref{f.16}) defines a unitary representation,
${U}_\otimes(\Lambda ,a)$, of the Poincar\'e group 
that satisfies the strong form of cluster properties because 
\beq
A \to A_{(jk)(i)}
\label{f.17}
\eeq 
when the interactions between particle $i$ and the other two 
particles are turned off.  Thus 
\[
U_\otimes (\Lambda ,a) \to 
A^{\dagger}_{(jk)(i)} {U}_{(jk)(i)}(\Lambda ,a) A_{(jk)(i)} =
\]
\[
A^{\dagger}_{(jk)(i)} A_{(jk)(i)} {U}_{(jk)}(\Lambda ,a)\otimes
{U}_{i}(\Lambda ,a) A_{(jk)(i)}^{\dagger}
A_{(jk)(i)} =
\]
\beq
{U}_{(jk)}(\Lambda ,a)\otimes
{U}_{i}(\Lambda ,a)
\label{f.18}
\eeq
This property ensures that the infinitesimal generators have the 
additive form (\ref{f.1}-\ref{f.4}) and (\ref{f.16})
generates the required three-body interactions.  

Because there are many other ways to construct symmetric products of
non-commuting operators and because it is possible to add a three-body
interaction to ${M}$ that commutes with and is independent of the total
momentum and spin, it is clear the three-body parts of the generators
that are required to restore the commutation relations are not unique.
It is also important to note that it is not possible to use the freedom
to add three-body interactions to eliminate the three-body
interactions required to restore the commutation relations; in this
representation the generated three-body interactions do not commute
with the non-interacting spin.
  
This construction can be extended to formulate dynamical models
satisfying the strong form of cluster properties for any fixed number
of particles, isobar models in any of Dirac's form of dynamics.  It is
even possible to treat production beyond isobar types of models.

Models with different kinematic subgroups can be constructed by
starting with different irreducible bases (\ref{c.2}), (\ref{c.3}).
As long as the reduced kernels of the interactions are identical, all
of the Bakamjian-Thomas three-body mass operators, $M$, will give
identical bound-state and scattering observables.  They are related 
by scattering equivalences constructed by applying the 
unitary transformations 
\beq
\vert (\lambda,j), \mathbf{v} , \mu, \cdots  \rangle = 
\vert (\lambda ,j), \mathbf{p}(\mathbf{v},\lambda )  , \mu , \cdots 
\rangle \lambda ^{3/2},
\label{f.19}
\eeq
or 
\beq 
\vert (\lambda,j), p^+ , \mathbf{p}_{\perp} , \mu,\cdots  \rangle = 
\vert (\lambda ,j), \mathbf{p}(p^+,\mathbf{p}_{\perp},\lambda )  , \mu',\cdots  \rangle 
\sqrt{{\omega_\lambda (\mathbf{p})  \over p^+ }}
D^j_{\mu' \mu} [B_c^{-1}(p) B_f(p)]
\label{f.20}
\eeq
on each invariant subspace of the associated mass operator.
Each of these representation is in turn scattering equivalent 
to a representation that satisfies strong cluster properties and 
has the same kinematic subgroup.

Because $A$ is a scattering equivalence, it is only necessary to solve
the Faddeev equations for the mass operator ${M}$ in a non-interacting
irreducible basis.  Furthermore, since all bound state and scattering
observables can be computed using only the internal mass operator,
with the delta functions in $\mathbf{p}$ and $\mu$ removed, this
equation is the same in all of Dirac's forms of dynamics when
expressed in terms of the kinematic mass and kinematically invariant
degeneracy quantum numbers.  The operators $A$ and the choice of
kinematic subgroup are only needed if the three-body system is
embedded in the four-body Hilbert space or if the eigenstates are used
to construct electroweak current matrix elements.

\section{Connection with quantum field theory}

Poincar\'e invariant quantum mechanics as formulated by Bakamjian and
Thomas resembles non-relativistic quantum mechanics more than quantum
field theory.  The Hilbert spaces have the same structure as
non-relativistic Hilbert spaces, the theory is not manifestly
covariant, spin $1/2$ particles are treated using two-component
spinors.  In spite of these apparent differences there is a direct
connection to quantum field theory which we outline below.

To develop the connection we assume the existence of an underlying
quantum field theory with a Poincar\'e invariant vacuum and a
collection of Heisenberg fields, $\bphi_i(x)$, where the bold face
indicates a multi-component field.  The index $i$ distinguishes different 
types of fields.
 
In quantum field theory Hilbert-space vectors are constructed by
applying functions of smeared fields,
\beq
\phi_i(f) = \int d^4 x \mathbf{f}(x) \cdot \bphi_i (x)
\label{g.1}
\eeq  
to the physical vacuum $\vert 0 \rangle$. 

Polynomials in the smeared fields applied to the physical vacuum
generate a dense set of vectors.
The field theoretic unitary representation of the Poincar\'e 
group $U^{\dagger} (\Lambda ,a)$ acts covariantly on the smeared fields:
\beq
U^{\dagger} (\Lambda ,a) \bphi_i(\mathbf{f}) U(\Lambda ,a)  
= \int d^4 x \mathbf{f}(\Lambda x+a )S(\Lambda)\bphi_i (x)
\label{g.3}
\eeq  
where $S(\Lambda)$ is the finite dimensional representation of the
Lorentz group appropriate to the field.  The covariance of the fields
implies Poincar\'e transformation properties of test functions that
leave the scalar product invariant.

If the field theory has one-particle states, then there are functions,
$A$, of smeared fields with the property that $A\vert 0 \rangle$ is a
one particle state.  One-particle eigenstates that transform
irreducibly with respect to the Poincar\'e group can be constructed by
projecting $A\vert 0 \rangle$ on states of sharp linear momentum and
canonical spin.  This can be done using the unitary representation
(\ref{g.3}) of the Poincar\'e group
\[
\vert (m,j) \mathbf{p}, \mu \rangle = A (\mathbf{p},\mu) 
\vert 0 \rangle :=  
\]
\beq
\sum_{\nu=-j}^j\int dR dp^0 d^4x  e^{i {p} \cdot {x}}U(R, {x} )
A \vert 0 \rangle  D^{j*}_{\mu \nu} (R) \delta (p^2 + m^2) \theta (P^0)  
\label{g.5}
\eeq
where $R$ is a rotation, $dR$ is the $SU(2)$ Haar measure, 
$U(R, {x} )$ is the unitary representation of the Poincar\'e group
restricted to rotations and spacetime translations, and
$D^{j*}_{\mu \mu} (R)$ is a $SU(2)$ Wigner function.

The normalization of these states can be chosen so 
\beq
\langle (m',j') \mathbf{p}', \mu' \vert (m,j) \mathbf{p}, \mu \rangle = 
\delta (\mathbf{p}'-\mathbf{p} ) 
\delta_{m'm} \delta_{j'j}
\delta_{\mu' \mu} .
\label{g.16}
\eeq
It follows from the definitions 
and the group representation properties that
these states transform as mass $m$ spin $j$ irreducible
representations of the Poincar\'e group:
\[
U(\Lambda ,a) \vert (m,j) \mathbf{p}, \mu \rangle =
\sum_{\mu'=-j}^j \vert (m,j) \bLambda{p}, \mu' \rangle e^{i\Lambda p \cdot a}
D^{j*}_{\mu' \mu} [B^{-1}(\Lambda (p)) \Lambda B(p)]
\sqrt{{\omega_m (\bLambda p) \over \omega_m (\mathbf{p})}}
  =
\]
\beq
\sum_{\mu'=-j}^j \int d\mathbf{p}'
\vert (m,j) \mathbf{p}', \mu' \rangle
{\cal D'}^{mj}_{\mu', \mathbf{p}';\mathbf{p},\mu}[\Lambda ,a]
\label{g.7}
\eeq
To construct scattering states define
$C(\mathbf{p},\mu):=  
(\sqrt{m^2 +\mathbf{p}^2}A(\mathbf{p},\mu))-[H,A(\mathbf{p},\mu)]_- )$.
Scattering states are then given by the Haag-Ruelle method:
\cite{Haag:1958vt}\cite{ruelle1}
\beq
\vert (\mathbf{p}_1, \mu_1, \cdots ,\mathbf{p}_N, \mu_N)^{\pm} \rangle = 
\lim_{t \to \pm \infty} U(-t) 
\prod_j [C_j (\mathbf{p}_j,\mu_j) e^{-i t\omega_{m_j} (\mathbf{p}_j)}]
\vert 0 \rangle   
\label{g.8}
\eeq
where the limits are strong limits after smearing over suitable 
momentum wave packets. 

The operators $\prod_j [C_j (\mathbf{p}_j,\mu_j)\vert 0 \rangle$
can be considered as mappings from an $N$
particle channel Hilbert space, ${\cal H}_{\alpha}$, to 
the Hilbert space of the field
theory.  Vectors in the $N$-particle channel Hilbert space are square
integrable functions in the variables $\mathbf{p}_1, \mu_1, \cdots
,\mathbf{p}_N, \mu_N$.  We denote these operators by $\Omega_{\alpha
  \pm}$ where $\alpha$ indicates the channel.

The direct sum of all of the channel Hilbert spaces, including the 
one-particle channels, defines an asymptotic Hilbert space. 
We define $\Omega_{\pm}$ that maps the asymptotic Hilbert space
to the physical Hilbert space by  
\beq
\Omega_{\pm} 
\left (
\begin{array}{l}
\vert \mathbf{f}_{\alpha_1} \rangle \\
\vert \mathbf{f}_{\alpha_2} \rangle \\
\vdots 
\end{array} 
\right ) =
\sum_\alpha \Omega_{\alpha_i \pm} \vert \mathbf{f}_{\alpha_i} \rangle   .
\label{g.9}
\eeq
By construction these wave operators satisfy the intertwining relations  
\cite{ruelle1}
\beq 
U(\Lambda ,a)  \Omega_{\pm} = \Omega_{\pm} \oplus_\alpha U_\alpha (\Lambda ,a).
\label{g.10}
\eeq
The Poincar\'e invariant $S$ operator of the field theory is given by 
\label{g.11}
\beq
S = \Omega_{+}^{\dagger}  \Omega_{-}
\label{g.12}
\eeq
where each $U_\alpha (\Lambda ,a)$ is a tensor product of single particle 
irreducible representations of the Poincar\'e group on the channel subspace 
${\cal H}_{\alpha}$

Poincar\'e invariant quantum mechanics formulated in the previous sections
has the same basic structure.  The primary difference is that the 
asymptotic Hilbert space for the field theory has an infinite number of 
channels and 
describes physics at all energy scales, while the Poincar\'e invariant 
quantum mechanical wave operators involve only a subset of these
channels that are experimentally relevant only up to a given energy scale.
  
If $\Pi$ is a Poincar\'e invariant projection operator on the asymptotic 
subspaces corresponding  channels of a Poincar\'e invariant quantum model 
that also limits the maximum invariant mass of the asymptotic states, then 
the following operator 
\label{g.13}
\beq
W= \Omega_{f+} \Pi \Omega^{\dagger}_{qm+} = 
\Omega_{f-} \Pi \Omega^{\dagger}_{qm-}  
\label{g.14}
\eeq
maps an invariant subspace of the quantum mechanical Hilbert space 
to an invariant subspace of the field theory Hilbert space in a manner
that satisfies
\beq
\Pi S_{qm} \Pi = \Pi S_{f} \Pi
\label{g.15}
\eeq
\beq
W U_p (\Lambda ,a) = U_f (\Lambda ,a) W ~.
\label{g.16a}
\eeq
These mappings define the relevant relation between the Poincar\'e
invariant quantum theory and the underlying field theory.


Thus, for asymptotic scattering states in the range of $\Pi$ the
Poincar\'e invariant quantum mechanical theory can be designed to 
give identical results to the field theory.  Obviously the two 
theories differ on asymptotic states that are not in the range of $\Pi$. 

Even though the Poincar\'e invariant quantum theory does not satisfy
microscopic locality, we see that it can give the same $S$ matrix
elements as the full field theory at a given energy scale.  

\section{Few nucleon applications}

In this section we discuss an illustrative set of applications to few
nucleon problems.  A realistic nucleon-nucleon interaction is needed
for these applications.  The invariant mass
operator for two free nucleons can be expressed in terms of a relative
momentum as
\beq
m_{012} =:\sqrt{{k}^2 +m_1^2} +
\sqrt{{k}^2 +m_2^2} .  
\label{h.1}
\eeq
It is always possible to express the two-body interaction as an addition 
to ${k}^2$:
\beq
{M}_{12}= m_{12} +{v}_{12} :=
\sqrt{{k}^2 + 2 \mu {v}_{nn} +m_1^2} +
\sqrt{{k}^2 + 2 \mu {v}_{nn} +m_2^2}  
\label{h.2}
\eeq
where following \cite{Coester:1975zz}
${v}_{nn}$ is a realistic nucleon-nucleon interaction
\cite{Wiringa:1994wb,Machleidt:2000ge}
and $\mu$ is the reduced mass
\beq
\mu = {m_1 m_2 \over m_1 + m_2}.
\label{h.3}
\eeq
In this representation the dynamical two-body 
mass operator becomes a function of
the non-relativistic center of mass Hamiltonian:
\beq
{M}_{12}=\sqrt{2 \mu { {h}_{nr}} +m_1^2} +
\sqrt{2 \mu { {h}_{nr}} +m_2^2}  
\label{h.4}
\eeq
where 
\beq
{h}_{nr} ={\mathbf{k}^2 \over 2 \mu} + {v}_{nr} .
\label{h.5}
\eeq
It is a consequence of the Kato-Birman 
\cite{kato:1966,Chandler:1976,baumgartl:1983} invariance principle 
that the relativistic wave
operators for (\ref{h.2}) and non-relativistic wave operators for
(\ref{h.5}) are identical
\[
{\Omega}_{nr \pm}:=  
\lim_{t \to \pm \infty}e^{i{H}_{nr}t } e^{-i H_{nr0} t } =
\lim_{t \to \pm \infty} e^{i{h}_{nr}t } e^{-i h_0 t } = 
\lim_{t \to \pm \infty} e^{i{M}t } e^{-i M_0 t } =
\]
\beq
\lim_{t \to \pm \infty} e^{i{M}^2t } e^{-i M^2_0 t } =
\lim_{t \to \pm \infty} e^{i{H}_r^2t } e^{-H^2_{r0} t } =
\lim_{t \to \pm \infty} e^{i{H}_rt } e^{-i H_{r0} t }  =
{\Omega}_{r \pm}
\label{h.6}
\eeq
where $M=M_{12}$ in (\ref{h.6}).
The identity (\ref{h.6}) ensures that both scattering operators are 
identical as functions of $\mathbf{k}^2$:
\beq
S_{nr}  = {\Omega}^{\dagger}_{nr +}{\Omega}_{nr -} =
{\Omega}^{\dagger}_{r +}{\Omega}_{r -} =
S_{r} 
\label{h.7} 
\eeq
Here the relativistic and non-relativistic $S$ are related because the
interactions are fit to the same two-body data correctly transformed
to the center of momentum frame.  The non-relativistic
Hamiltonian (\ref{h.5}) is NOT the non-relativistic limit of
(\ref{h.2}).

This construction, which first appeared in \cite{Coester:1975zz}, 
shows that existing realistic interactions can be directly used in the
formulation of a Poincar\'e invariant two-body problem.  Equation
(\ref{h.4}) implies that the wave functions of (\ref{h.5}) and
(\ref{h.2}) are identical functions of $\mathbf{k}^2,l,s$.

If we replace the interaction in (\ref{d.6}) by the interaction 
$v_{12}$ in (\ref{h.2}) and use this in the three-body calculation 
discussed in section \ref{sec6} then the three-body $S$-matrix can be 
expressed in terms of three-body mass operators:
\beq
\bar{S}_{ac} = \delta_{ac} - 2 \pi i \delta(M_a-M_c) {T}^{ac}(z_c)  
\label{h.8} 
\eeq
which are functions of the transition operators
\beq
{T}^{ac} (z) = {T}^{ac}_{}(z)  = {V}_{}^c + 
{V}^a {R}(z) {V}^c 
\label{h.9} 
\eeq
where $a,b,c \in  \{ (12)(3), (23)(1) , (31)(2)\}$,
\beq
{M}_{(ij)(k)}=\sqrt{ 
(\sqrt{m_i^2 + k^2 +2 \mu {v}_{nn} } + 
\sqrt{m_j^2 + k^2 +2 \mu {v}_{nn} })^2
+q^2} + \sqrt{ {m}_{k}^2+q^2} .  
\label{h.9a}
\eeq
\beq
{V}_{a}={M}_{a} - M_0 
\qquad 
{V}^c = \sum_{a \not=b} {V}_{a} 
\qquad
{R}(z) = (z -{M})^{-1}
\label{h.10} 
\eeq
\beq
{R}_{c}(z) = (z -M_0-{V}_{c})^{-1}  
\label{h.12} 
\eeq
\beq
{R}(z) = {R}_{c} (z) + {R}_{c} (z) {V}^c {R}(z) 
\label{h.13} 
\eeq
and ${v}_{nn}$ is the nucleon-nucleon interaction that appears in
(\ref{h.2}) embedded in the three-nucleon Hilbert space with the delta
functions in (\ref{f.8}).  The Faddeev equations can be derived using
standard methods
\beq
{T}^{ab}(z)  =
{V}^b + \sum_{c\not= a} {V}_{c}  {R}_{c} (z) {T}^{cb} (z) .
\label{h.14} 
\eeq
While it does not make any sense to study the non-relativistic limit of
interactions that are constructed by fitting to two-body bound and 
scattering data, we can compare the relativistic and non-relativistic 
three-body calculations that use the same two-body interaction, $v_{nn}$,
as input.  In the Poincar\'e invariant quantum mechanics case 
the Faddeev equations have the form 
\beq
\langle a \vert {T}^{ab}(z)\vert b \rangle  =
\langle a \vert  {V}^b \vert b \rangle
 + \sum_{c\not= a}\int  \langle a \vert c \rangle 
\langle c \vert  {V}_{c}  {R}_{c} (z)\vert c \rangle 
\langle c \vert  {T}^{cb} (z) \vert b \rangle
\label{h.15} 
\eeq
where
$
\langle a \vert c \rangle$ are Poincar\'e group Racah coefficients,
which are the unitary transformation that relate three-body 
Poincar\'e irreducible bases constructed using pairwise coupling in 
different orders.  These coefficients, 
which have the form
\beq
\langle a \vert c \rangle = \delta (\mathbf{p}-\mathbf{p'})
\delta_{\mu \mu'}
\delta (m-m') \delta_{jj'} 
{\cal R}^{mj} (d_a, d_c),
\label{h.16} 
\eeq
with $d_a$ and $d_b$ distinct sets of invariant degeneracy 
parameter, replace the non-relativistic permutation operators.

The construction of the kernel is facilitated by the fact that the 
two-body eigenfunctions of (\ref{h.4}) and (\ref{h.5}) are identical.
The kernel of the relativistic Faddeev equation can be directly related 
to the non-relativistic two-body $t$ using the following relations:
\[
{ \langle c' \vert {V}_{c} {R}_{c} (z_c) \vert c \rangle} =
\langle c' \vert {T}_{c}(z_c)  (z_c-{M}_0)^{-1} \vert c \rangle =
\]
\[
\langle c' \vert {V}_{c} \vert c^- \rangle (z_c-M_0)^{-1} =  
\langle c' \vert {M}_{c} -M_0  \vert c^- \rangle (z_c-M_0)^{-1} =
\]
\beq
{2 \mu \over \omega_1 \omega_2 + \omega_1'\omega_2'}
{(\omega_1 + \omega_2)^2 + (\omega'_1 + \omega'_2)^2 
\over 
\sqrt{
(\omega_1 + \omega_2)^2 +q^2}+ 
\sqrt{
(\omega'_1 + \omega'_2)^2 + q^2} 
}
\langle c'\vert t_{nr}(k_c) \vert 
c \rangle 
(z_c-M_0)^{-1}
\label{h.21}
\eeq
where
\beq
\omega_i = \sqrt{{k}^2 + m_i^2}~, 
\label{h.22}
\eeq
which holds for the half shell kernel;  this can be used as input 
to construct the fully off-shell kernel using the first 
resolvent identity \cite{Keister:2005eq} 
\beq
{T}_c(z') = {T}_c(z_c) + {T}_c(z') 
{ (z'-z_c)\over
(z'-M_0)(z_c-M_0)} {T}_c (z_c) 
\qquad z' \not=z_c
\label{h.17} 
\eeq
Alternatively, this kernel has also been computed using an iterative 
procedure based on a non-linear integral equation\cite {Kamada:2007ms}.

The differences with the non-relativistic three-body calculations are
the different off-shell dependence dictated by (\ref{h.21}), the
differences in the Poincar\'e group Racah coefficients (\ref{h.16})
and the non-relativistic permutation operators.  These differences
show up for the first time in the three-body system, since our
two-body interactions are designed to reproduce the same experimental
two-body cross sections as the non-relativistic calculations.

Solving these equations leads to three types of predictions: 
binding energies,
\beq
M  \vert \Psi \rangle = {\lambda} \vert \Psi \rangle 
\label{h.19} 
\eeq
\beq
\vert \Psi \rangle = E(V) \vert {\Psi}_{bt} \rangle
\qquad 
{M}_{bt} \vert {\Psi}_{bt} \rangle = {\lambda} 
\vert {\Psi}_{bt} \rangle , 
\label{h.20} 
\eeq
scattering probabilities ($N=3$ only),
\beq
\vert S_{fi} \vert^2  = 
\vert \langle {\Psi}_{f}^+ \vert {\Psi}_{i}^- \rangle \vert^2= 
\vert \langle {\Psi}_{btf}^+ \vert  {\Psi}_{bti}^- \rangle\vert^2 ~,
\label{h.21a} 
\eeq
electromagnetic and weak current matrix elements
\beq
\langle {\Psi}_{f} \vert I^{\nu} (0) \vert{\Psi}_{i} \rangle 
= 
\langle {\Psi}_{f} \vert {A}
I^{\nu} (0) {A}^{\dagger} \vert {\Psi}_{i} \rangle ~,
\label{h.22a} 
\eeq
where $I^{\mu}(0)$ is a current that is conserved, 
covariant, and clusters in the representation (\ref{f.16}) 
of the three-body dynamics.

In what follows we discuss three applications of Poincar\'e invariant 
quantum mechanics that illustrate its ability to model 
a variety reactions where relativity may be important.

\subsection{Relativisitic spin rotations in low energy $A_y$} 

A calculation by Miller and Schewnk \cite{Miller:2007ad} suggested
that Wigner rotations might have an observable effect on the
polarization observable $A_y$ for low-energy p-d scattering.
Comparison of three-body calculations based on Poincar\'e invariant
quantum mechanics \cite{Witala:2008va} and non-relativistic quantum
mechanics using the same realistic CD-Bonn interaction
\cite{Machleidt:2000ge} as input indeed show a surprising sensitivity
of $A_y$ to Wigner rotations.  These calculations, which are shown in
Fig. 8.1, compare the non-relativistic result (dotted curve), the
relativistic result without Wigner rotations (dashed curve) and the
full relativistic calculation (solid curve) to 
data \cite{cub:1989,Tornow:1991}.  While the relativistic effects move the
calculations away from the data, this calculation illustrates that the
relativistic effects cannot be ignored in these calculations, even at
these low energies.

\subsection{Relativistic Effects in Exclusive pd Breakup} 

The value of Poincar\'e invariant quantum mechanics is that it
provides a consistent framework to study strong interactions in the
few GeV energy scale.  At this scale it is more efficient to perform
calculations using direct integration
\cite{Fachruddin:2000wv,Liu:2004tv} rather than with partial
wave expansions.  The feasibility of using Poinca\r'e invariant quantum
mechanics to treat nucleon deuteron-scattering at these energy scales
was established by solving the Faddeev equation of section \ref{sec6} using
Malfliet-Tjon \cite{Malfliet:1968tj} interactions to model 
the nucleon-nucleon potential.
The two-body interactions were included using the method discussed
above.  Convergence of the solutions of the Faddeev equations was
established up to 2GeV
\cite{Lin:2007ck,Lin:2007kg,Lin:2008sy}.
In three-body reactions there are many
observables that can be used to test the sensitivity of relativistic
effects.  One interesting observable is the cross section when 
the outgoing protons in a breakup reaction are measured at
symmetric angles relative to the beam direction.  These cross
sections were computed \cite{Lin:2007kg} 
in non-relativistic and Poincar\'e invariant
three-body models using the same Malfliet-Tjon two-body interactions
as input. 

Fig. 8.3 shows cross sections for different choices of angles
symmetric about the beam direction.  The solid curve is the
relativistic calculation while the long dashed curve is the
non-relativistic one.  The other two curves compare the exact
calculation to the first terms in the multiple scattering series both
for the relativistic and non-relativistic cases.  As
the angle is increased the relativistic and non-relativistic curves,
exhibit different behavior.  For this kinematic configuration
the multiple scattering series converges quickly, although this result
depends on what is measured.  Fig. 8.2 shows similar plots for
non-symmetric angles.  Again the first order multiple scattering
calculations work reasonably well and there is a definite difference
between the relativistic and non-relativistic predictions.  In both
cases the data \cite{Punjabi:1988hn}, has the same qualitative 
behavior as the relativistic calculations.

\subsection{Exchange currents in electron-deuteron scattering}

The last application involves electron scattering off of nuclear targets
at values of momentum transfer $Q^2$ appropriate to J-lab experiments.  
In Poincar\'e invariant 
quantum mechanics electron scattering observables in the one-photon-exchange 
approximation can be expressed in terms of matrix elements of a 
conserved covariant current 
$I^{\mu} (x)$ which should have a cluster expansion 
\beq
I^{\mu} (x) = \sum I_i^{\mu} (x) + \sum I_{ij} ^{\mu} (x)
+ \sum I_{ijk}^{\mu} (x) + \cdots .
\eeq
Both Poincar\'e covariance, current conservation, and cluster
properties put dynamical constraints on the current operator.

The deuteron is the simplest electromgnetic target that is sensitive
to the two-body part of the current.  While a general method for
constructing $I^{\mu} (x)$ based on dynamical considerations is not
known, the constraints can be satisfied by using the Wigner-Eckart
theorem for the Poincar\'e group, which amounts to computing a maximal
set of linearly independent current matrix elements and using
covariance and current conservation to generate the remaining matrix
elements.  Different model two-body currents can be tested in this
framework. For elastic scattering off of a deuteron there are three
independent observables which can be taken as, $A(Q^2)$, $B(Q^2)$, and
 $T_{20}(Q^2, 70^o)$.  The input to a calculation is a deuteron wave
function, a dynamical representation of the Poincar\'e group, nucleon
form factors
\cite{nnformfactor-Lomon1,nnff-BBA,nnff-BBBA,nff-kelly,nnformfactor-BI}, 
and a model exchange current
\cite{Riska:1989bh}.   The calculations
illustrated in Figs. 8.4-8.6 
use a model of the deuteron with a light front kinematic
symmetry.  The dynamical representation of the Poincar\'e group is
constructed from the Argonne V18 interaction\cite{Wiringa:1994wb},
and the exchange current
is the long-range part of a ``pair current'' derived from the 
one-pion-exchange part of the V18 interaction.  Figs. 8.4, 
8.5, and 8.6 show
comparisons of $A(Q^2)$, $B(Q^2)$, and $T_{20}(Q^2, 70^o)$ to experimental
data with and without the exchange current.  Two different implementations
of the Poincar\'e group Wigner-Eckart theorem are responsible for the small 
difference in the curves labeled II and III.  

These three calculations illustrate both the power and flexibility of 
Poinca\'e invariant quantum mechanics as a tool to study systems of 
strongly interacting particles at scales up to a 
few GeV.  Data shown for $A$ are from 
\cite{buchanan65,Elias69,Benak66,Arnold75,Platch90,Galster71,cramer85,simon81,JlabHallc99,JlabHalla99,Berard74},
for $B$ are from 
\cite{Bosted90,martin77,cramer85,auffret85,simon81,buchanan65},   
and for 
$T_{20}$  from \cite{Dmitr85,Novos86,Gilman90,Schul84,The91,Jlabt20}.

These calculations demonstrate that Poincar\'e invariant 
quantum mechanics is a useful framework for making realistic models
of system of strongly interacting particles at the few-GeV energy 
scale.   Some of these effects extend to surprisingly low energies. 
  
This work was performed under the auspices of the U.~S.  Department of
Energy, Office of Nuclear Physics, under contract
No. DE-FG02-86ER40286.




\begin{figure}
\begin{center}
\includegraphics[width=12.0cm,clip]{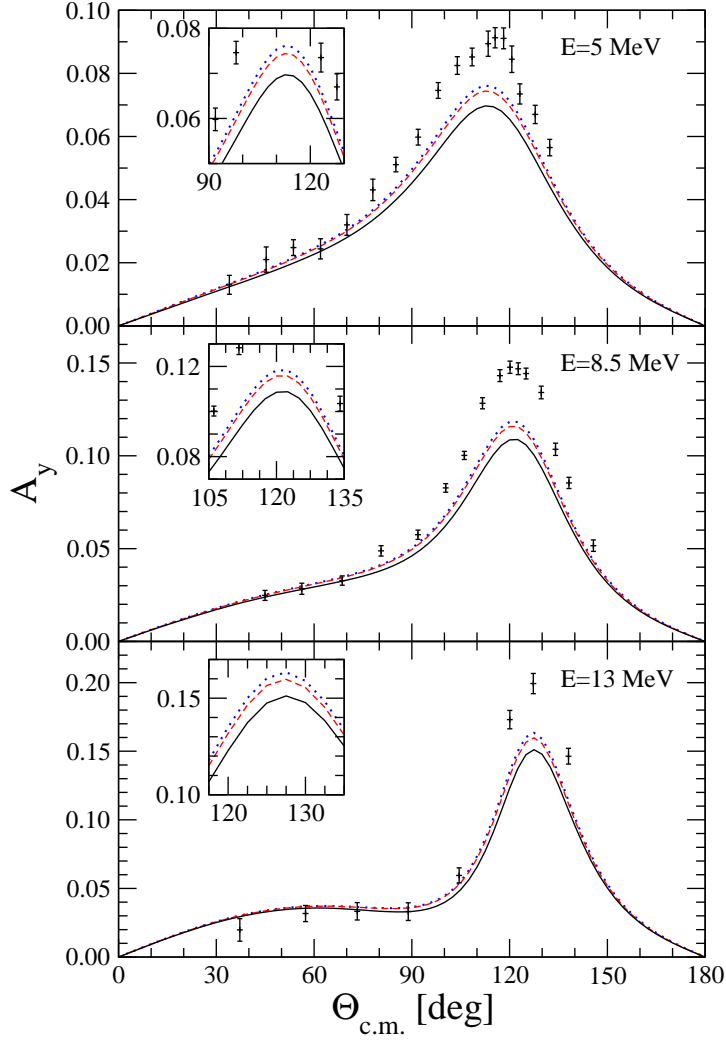}
\caption[Short caption for figure 1]{\label{labelFig1} Comparison of 
relativistic and non-relativistic calculations of the observable $A_y$
at low energies.}
\end{center}
\label{fig.1}
\end{figure}

\begin{figure}
\begin{center}
\includegraphics[width=12.0cm,clip,bb=60 28 300 500]{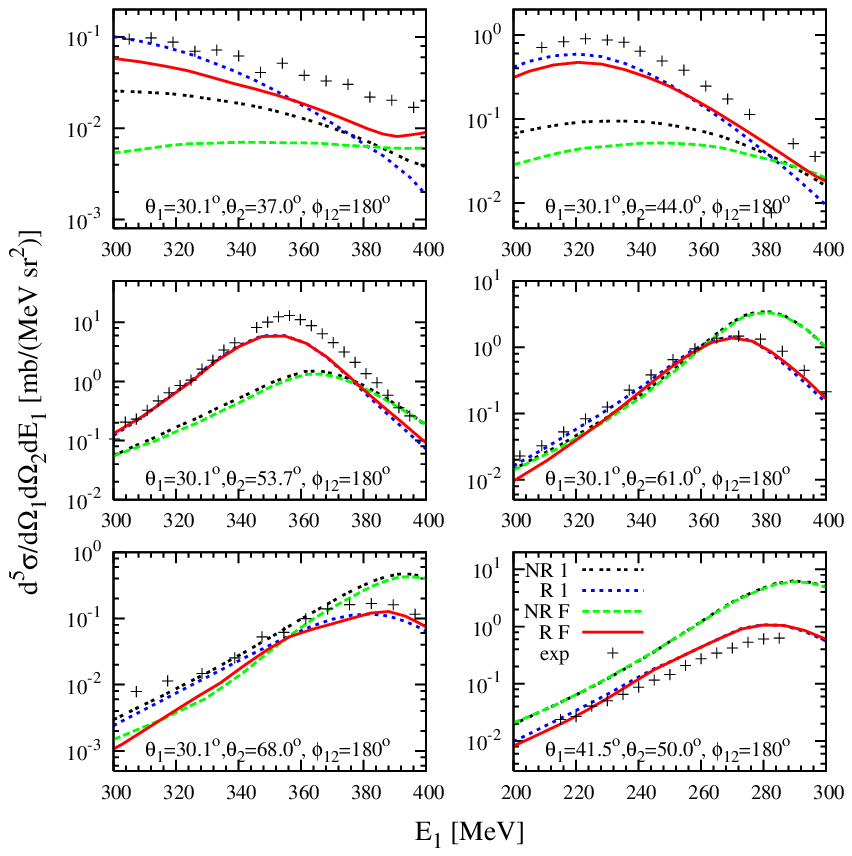}
\caption[Short caption for figure 2]{\label{labelFig2} Comparison of 
relativistic and non-relativistic calculations of exclusive proton 
deuteron breakup scattering at non-symmetric angles.}
\end{center}
\label{fig.3}
\end{figure}


\begin{figure}
\begin{center}
\includegraphics[width=12.0cm,clip]{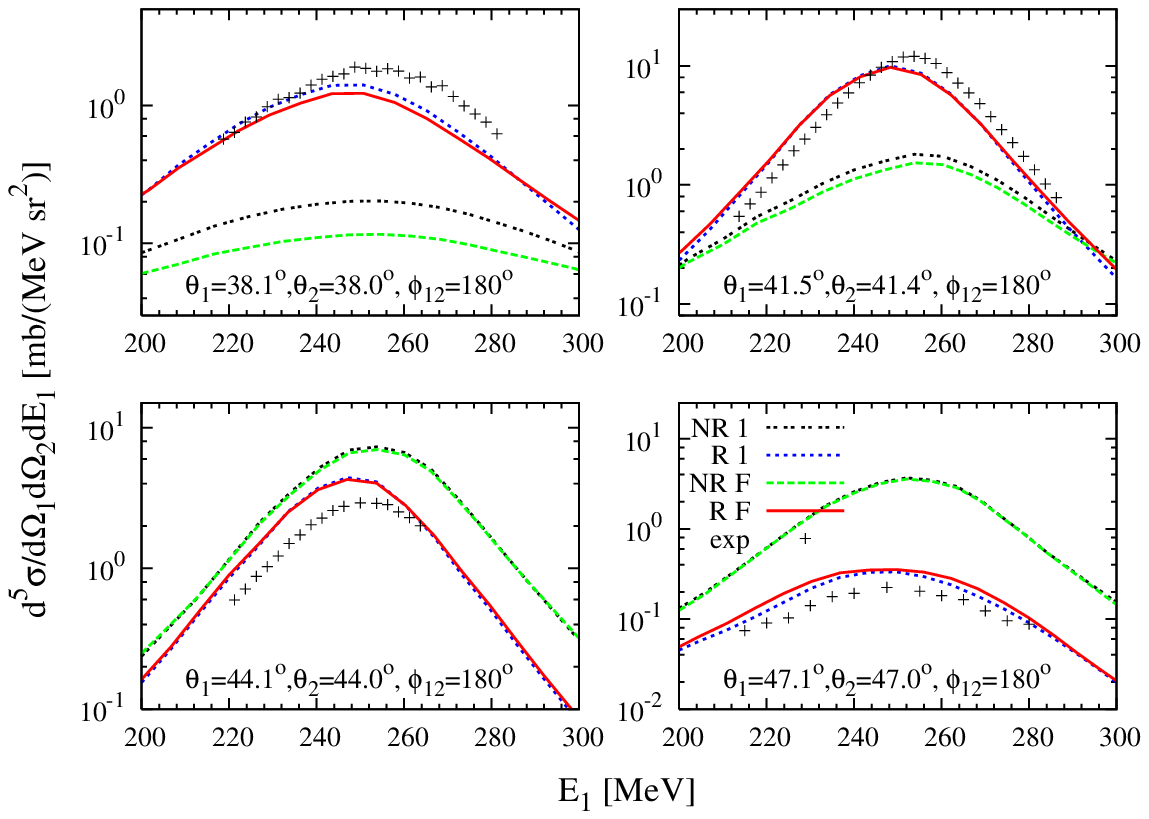}
\caption[Short caption for figure 3]{\label{labelFig3}
Comparison of 
relativistic and non-relativistic calculations of exclusive proton 
deuteron breakup scattering at non-symmetric angles.
}
\end{center}
\label{fig.2}
\end{figure}

\begin{figure}
\begin{center}
\includegraphics[width=12.0cm,clip]{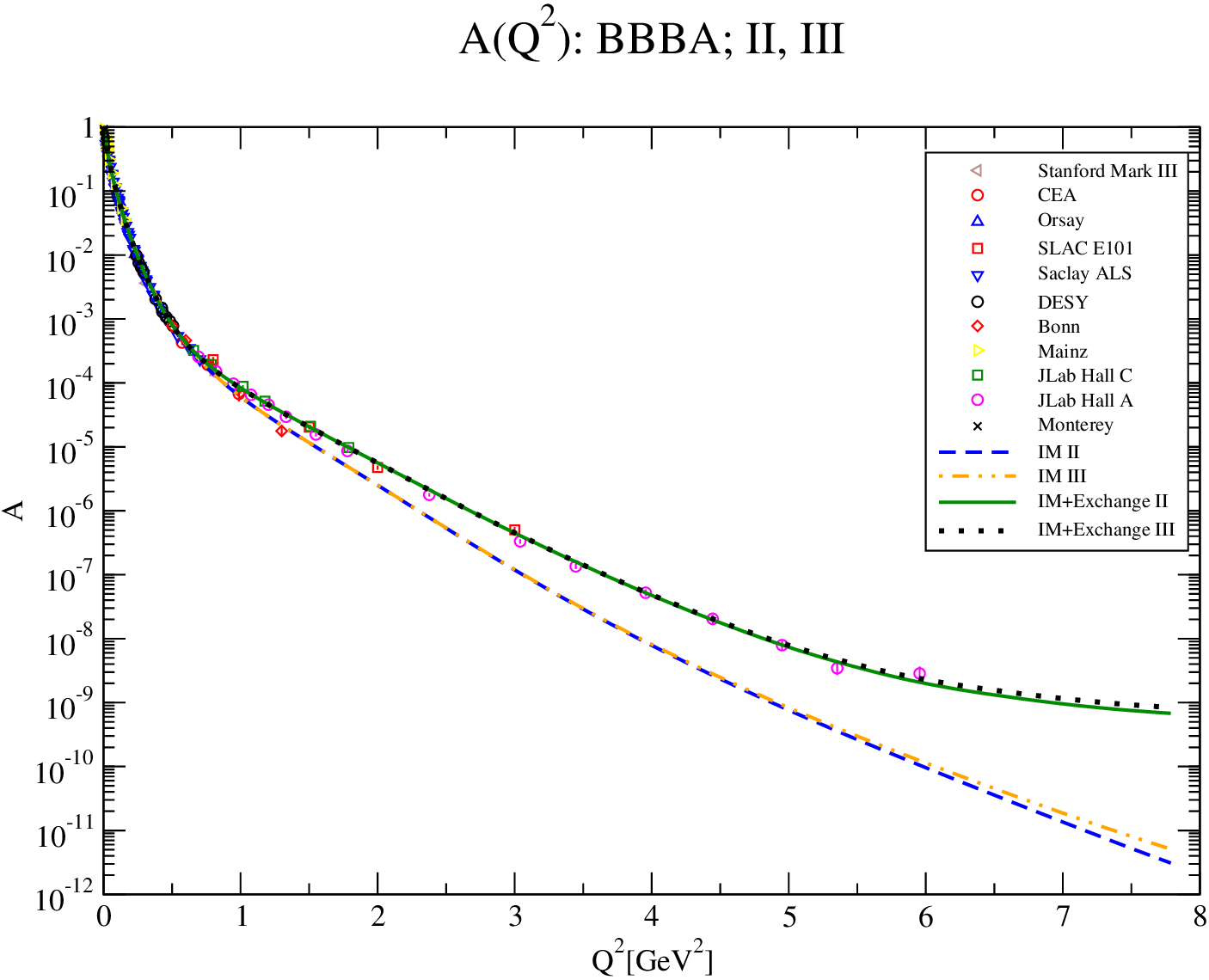}
\caption[Short caption for figure 4]{\label{labelFig4}
Comparison of
a front-form calculation of the elastic electron-deuteron scatting
observable $A(Q^2)$ with and without ``pair current'' contributions.
}
\end{center}
\label{fig.4}
\end{figure}

\begin{figure}
\begin{center}
\includegraphics[width=12.0cm,clip]{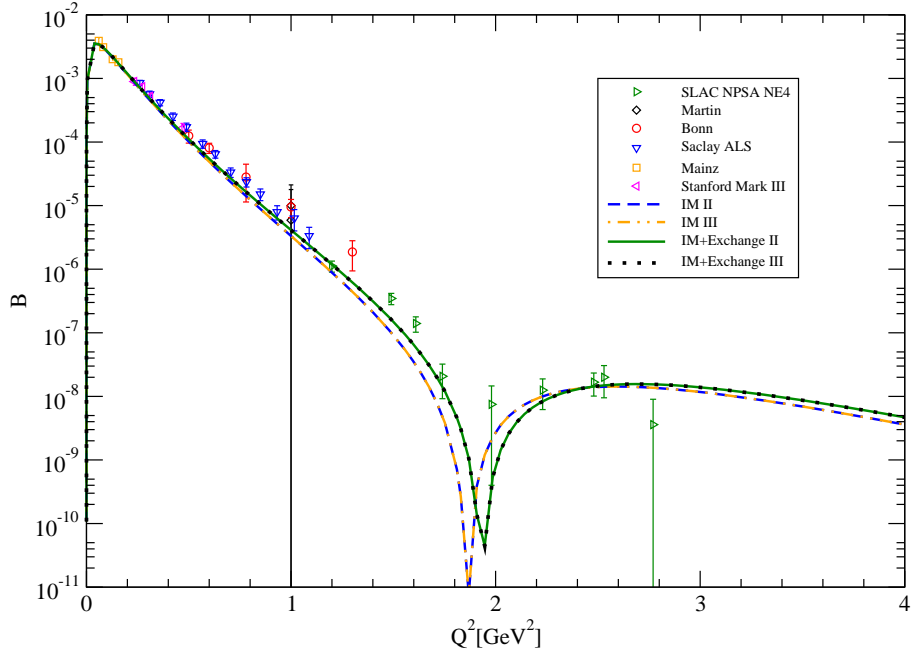}
\caption[Short caption for figure 5]{\label{labelFig5} Comparison of
a front-form calculation of the elastic electron-deuteron scatting
observable $B(Q^2)$ with and without ``pair current'' contributions.
}
\end{center}
\label{fig.5}
\end{figure}
\begin{figure}
\begin{center}
\includegraphics[width=12.0cm,clip]{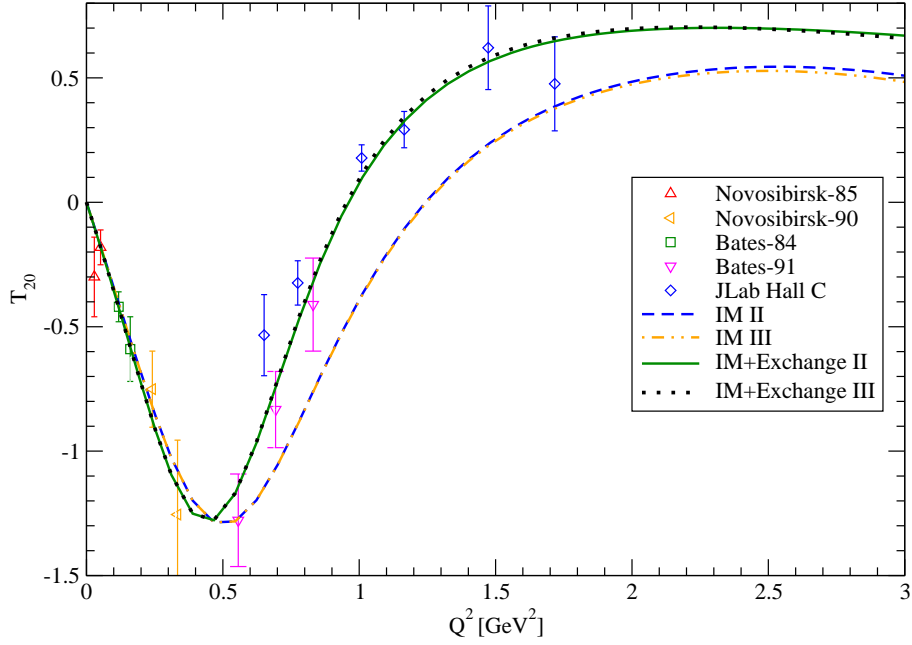}
\caption[Short caption for figure 6]{\label{labelFig6} 
Comparison of
a front-form calculation of the elastic electron-deuteron tensor polarization 
with and without ``pair current'' contributions.}
\end{center}
\label{fig.6}
\end{figure}


\end{document}